\documentclass{article}

\PassOptionsToPackage{numbers, compress}{natbib}

\usepackage[final]{neurips_2021}

\usepackage[utf8]{inputenc} % allow utf-8 input
\usepackage[T1]{fontenc}    % use 8-bit T1 fonts
\usepackage[hidelinks]{hyperref}       % hyperlinks
\usepackage{url}            % simple URL typesetting
\usepackage{booktabs}       % professional-quality tables
\usepackage{amsfonts}       % blackboard math symbols
\usepackage{nicefrac}       % compact symbols for 1/2, etc.
\usepackage{microtype}      % microtypography
\usepackage{xcolor}         % colors
\usepackage{graphicx}
\usepackage{caption}
\usepackage{subcaption}

\title{Characterizing Human Explanation Strategies\\to Inform the Design of Explainable AI\\for Building Damage Assessment}

\author{%
  Donghoon Shin\\
  Seoul National University \\
  Seoul, Republic of Korea \\
  \texttt{\href{mailto:ssshyhy@snu.ac.kr}{ssshyhy@snu.ac.kr}} \\
  \And
  Sachin Grover \\
  Carnegie Mellon University \\
  Pittsburgh, PA, USA \\
  \texttt{\href{mailto:sachingr@andrew.cmu.edu}{sachingr@andrew.cmu.edu}} \\
  \AND
  Kenneth Holstein \\
  Carnegie Mellon University \\
  Pittsburgh, PA, USA \\
  \texttt{\href{mailto:kjholste@cs.cmu.edu}{kjholste@cs.cmu.edu}} \\
  \And
  Adam Perer \\
  Carnegie Mellon University \\
  Pittsburgh, PA, USA \\
  \texttt{\href{mailto:adamperer@cmu.edu}{adamperer@cmu.edu}} \\
}

\begin{document}
\bibliographystyle{plainnat}

\maketitle

\begin{abstract}
Explainable AI (XAI) is a promising means of supporting human-AI collaborations for high-stakes visual detection tasks, such as damage detection tasks from satellite imageries, as fully-automated approaches are unlikely to be perfectly safe and reliable. However, most existing XAI techniques are not informed by the understandings of task-specific needs of humans for explanations. Thus, we took a first step toward understanding what forms of XAI humans require in damage detection tasks. We conducted an online crowdsourced study to understand how people explain their own assessments, when evaluating the severity of building damage based on satellite imagery. Through the study with 60 crowdworkers, we surfaced six major strategies that humans utilize to explain their visual damage assessments. We present implications of our findings for the design of XAI methods for such visual detection contexts, and discuss opportunities for future research.

\end{abstract}

\section{Introduction}

Public satellite image datasets of natural disasters, such as xBD~\cite{gupta2019xbd}, have been used to develop AI tools for assessing building damage from satellite imagery. However, these tools remain imperfect and error prone~\cite{xview_first}, and it is still not safe to fully rely on them in high-stakes contexts. Recent studies point to the potential of human-AI collaborative approaches to improve performance on damage assessment tasks~\cite{logar2020pulsesatellite, zhang2019crowdlearn}, and explainable AI (XAI) approaches are expected to play a major role in supporting effective human-AI collaborations in this and related domains~\cite{mohseni2021multidisciplinary}. Yet, most existing XAI techniques are not informed by an understanding of the types of explanations that are likely to be meaningful and informative to human decision-makers in particular tasks~\cite{arrieta2020explainable, miller2019explanation}.

As a first step toward understanding what humans require from XAI in building damage assessment tasks, we begin by characterizing how humans \textit{generate} explanations for their own assessments in such tasks. Specifically, we conducted an online study to collect data on how people explain their own assessments in the context of damage detection from satellite imagery. Using the web annotation system designed for our study, we ran an online crowdsourced study with 60 crowdworkers and asked them to annotate and assess the damage on 10 pairs of satellite imagery, each of which consists of pre- and post-disaster image. Through open-coding of participants' responses, we observed six major strategies that humans utilized to explain their assessments. Based on our findings, we discuss opportunities for future research.
\section{Application Context}

This study reports our preliminary findings on how humans explain their assessments of building damage caused by natural disasters from satellite imageries. As such, we forecast that our results would contribute to informing more human-centered explanations for designing XAI targeted to building damage assessment tasks.

To collect the explanations of building damage assessment from participants, we utilized satellite images from the xBD dataset~\cite{gupta2019xbd}, designed to support HADR research. xBD includes pre/post-disaster pairs of satellite imagery, showing building damage from various natural disasters such as floods, fires, earthquakes, and hurricanes. For each case, xBD offers a ground truth damage assessment (i.e., coordinates and extent of damage), which we use to provide feedback to participants during training.
\section{Method}

\subsection{Annotation system}

\begin{figure}[h!]
    \centering
    \includegraphics[width=.95\columnwidth]{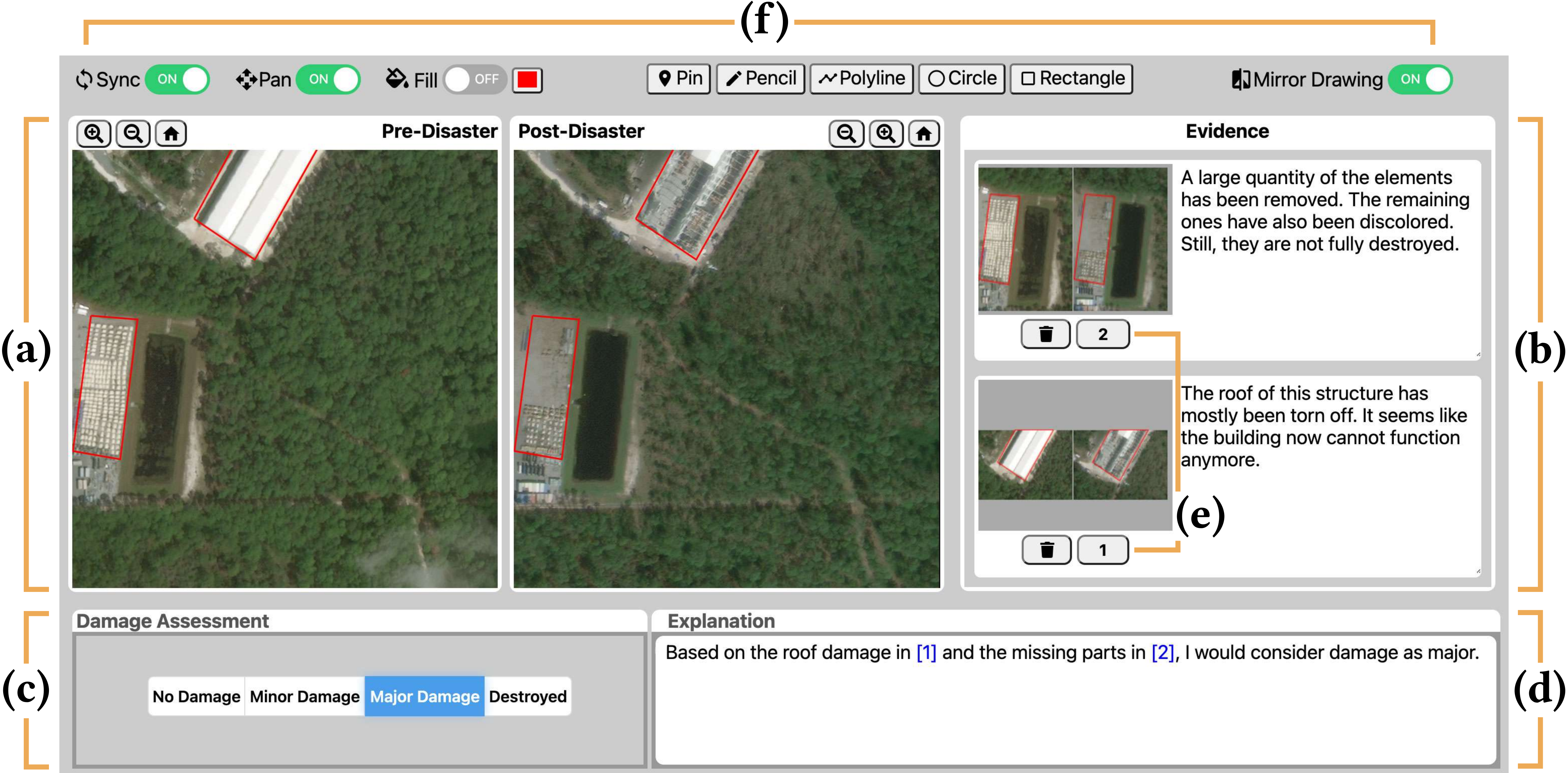}
    \caption{Screenshot of our experimental annotation system. (a) Side-by-side, interactive display of pre- and post-disaster images; (b) Editable text fields for user-generated image annotations; (c) Buttons for users to enter their overall assessment of building damage; (d) Text field for users to explain their overall assessment, possibly with linked references to specific annotations; (e) Buttons for adding direct references to specific image annotations, when writing an explanation; (f) Annotation tools to support users in marking up images}
    \label{fig:keyscreen}
\end{figure}

To collect human explanations of damage assessments, we developed a new, web-based annotation system as shown in Figure~\ref{fig:keyscreen}. In this system, users can view pre- and post-disaster images side-by-side, and can draw markups directly on each image (Figure~\ref{fig:keyscreen}-a) using various drawing tools (Figure~\ref{fig:keyscreen}-f). Users have the option to add notes to each markup they create, in an “Evidence” panel  (Figure~\ref{fig:keyscreen}-b). Finally, the annotator is then asked to check their assessment of the extent of damage (Figure~\ref{fig:keyscreen}-c) and provide an explanation for this assessment (Figure~\ref{fig:keyscreen}-d). Here, the annotator may reference specific annotations within their explanation (Figure~\ref{fig:keyscreen}-e).

\paragraph{Participants \& Study procedure.}
We recruited participants through Prolific, an online crowdsourcing platform, with English as their primary language. We received complete responses from a total of 60 participants (\textit{M$_{age}$} = 34.07, \textit{SD$_{age}$} = 8.28; 26 female). For the purposes of this study, participants were shown satellite images from the xBD dataset in which a \textit{single} building (potentially out of multiple buildings shown in the image) was damaged. As such, participants were required to assess both \textit{which} and \textit{to what extent} building was damaged.

Each participant took part in a training session before proceeding to the main session of the study. During this training phase, participants were shown a sequence of cases. For each case, participants were asked to make a damage assessment, but were not yet asked to explain their assessments. Upon making a damage assessment, the system provided immediate feedback on a participant's response, while also displaying the “ground truth” (an expert's assessment). 

To help participants learn both the damage assessment task and how to work with the annotation interface, the training session consisted of three consecutive parts, gradually decreasing the level of guidance provided. During the first part of the training, the system showed a spotlight over the location of the damaged building and automatically zoomed in on this location, so that participants could focus on learning how to assess damage without the additional overhead of searching an image for the damaged building. In the second part, although the spotlight was still shown, the system did not zoom in on the location of the damaged building. This provided participants with opportunities to begin making inferences based on the broader surroundings of a building, and perhaps to zoom and pan on their own as needed. Finally, the third phase of the training mirrored the main session of the study. In this phase, the system stopped spotlighting the damaged building, in order to provide participants opportunities to practice searching an image for the damaged building. Throughout this training phase, participants had access to a tooltip, which allowed the to review the provided criteria for assessing the level of damage at any time.

After completing the training session, participants then moved on to the main session. Each participant was asked to assess the level of building damage on a series of 10 cases, and to explain their assessment in each case, using the interface shown in Figure~\ref{fig:keyscreen}. The average time participants spent for the training and main session is 13.26 minutes (\textit{SD} = 5.52) and 23.71 minutes (\textit{SD} = 12.19), respectively.

\paragraph{Analysis.}
We used an iterative, open coding approach~\cite{walker2006grounded} to identify categories among the explanations that participants generated. First, two independent coders read through the overall responses of participants and generated initial high-level themes. The coders independently coded cases using these themes, and then came together to compare and discuss their coding, making edits to the codes accordingly. Finally, one coder re-coded the cases and examined the content and distribution of the resulting codes, as described in Section~\ref{result}.
\section{Result} \label{result}

\subsection{Codes}

\paragraph{A: Constructing a causal argument to explain building damage.}

In some cases, rather than directly referencing visual features of a building itself, participants instead pointed to visual evidence of a natural disaster in a building's surroundings to explain their assessment of building damage (A-1; e.g., \textit{“From the evidence of flooding, I would say the building seems to have been affected”}). In other cases, participants inferred that a particular type of natural disaster had occurred based on evidence of damage to a building, and then explained their overall assessment of building damage with reference to the type of disaster (A-2; e.g., \textit{“The building has roof damage. Probably a hurricane came and hit it”}). Some participants constructed more complex, multi-step causal arguments (A-3; e.g., \textit{“(Step 1) There was a fire and (Step 2) it was a wildfire that took everything from the building. (Step 3) You can only see the outline of the building”}).

\begin{figure}
    \centering
    \includegraphics[width=\textwidth]{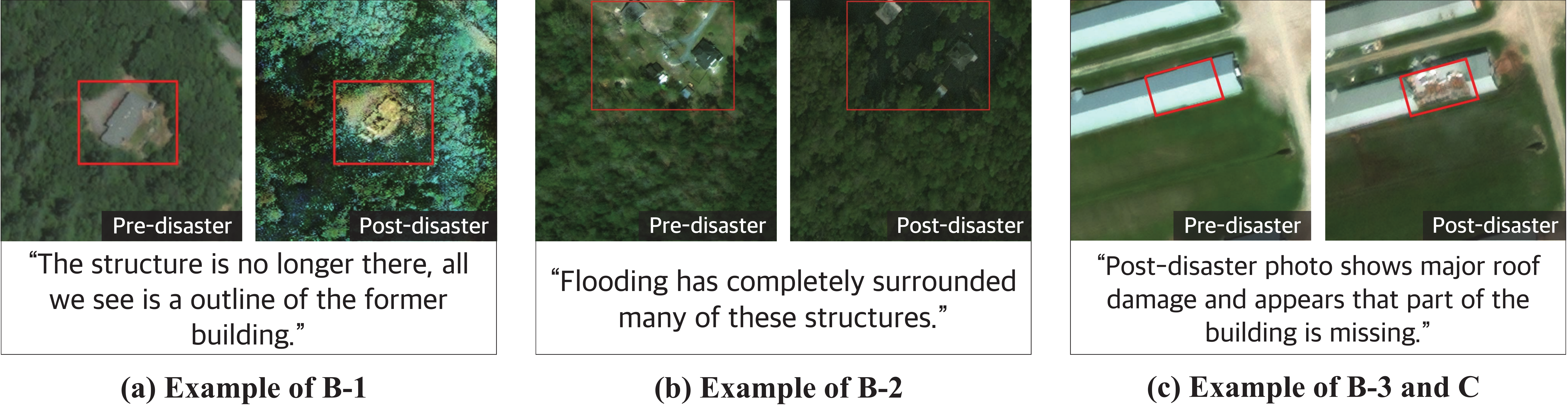}
    \caption{Example sets of markup and annotation that were coded as B-1, B-2, B-3, and C}
    \label{fig:examples}
\end{figure}

\paragraph{B: Contrasting pre- and post-disaster imagery.}
Many participants explained their damage assessments through direct comparisons across pre- and post-disaster images (e.g., by creating an annotation using the mirror-drawing tool). In more than 200 cases, participants referenced contrasts in the appearance of a specific building between the pre- and post-disaster images (B-1; see Figure~\ref{fig:examples}-a), or contrasts in the appearance of specific sub-structures of a building (B-3; see Figure~\ref{fig:examples}-c). In addition, participants often directly compared the pre- and post-disaster appearance of the area surrounding a building (B-2; see Figure~\ref{fig:examples}-b). Finally, ambiguous cases in which people generated contrast-based explanations, but did not clearly specify which elements of an image they were comparing, were marked with the code B-U.

\paragraph{C: Highlighting affected part of a building.}
Rather than drawing markup around the whole building, some participants referenced specific affected parts of the building, but without necessarily comparing pre- and post-images in their explanations (Figure~\ref{fig:examples}-c).

\paragraph{D: Explanations based on the extent of damage to a specific building.}
Some participants explained their assessment of the level of damage to a given building based on the \textit{proportion} of the building that appears to be damaged (D-1; e.g., \textit{“Approximately a half of the building was collapsed”}). Interestingly, even when significant building damage was evident, some participants explained lower assessments of damage by arguing that the damage appeared repairable (D-2; e.g., \textit{“One part (of the building) was hit ... seems like it could be rebuilt”}).

\paragraph{E: Explaining reasons for lack of confidence in their own assessment.}
In cases where participants felt under-confident in their damage assessment, they sometimes provided reasons for this under-confidence as part of their explanations. For example, some participants signaled their lack of confidence with reference to properties of a satellite image, such as low-resolution, visual distortion, or small buildings being obscured by shadows or taller buildings (E-1; e.g., \textit{“The imagery has great distortion and is difficult to judge”}). Other participants pointed out that they saw other changes between the pre- and post-disaster images (e.g., newly-built buildings in the area surrounding a given building), which made it challenging to precisely assess building damage (E-2; e.g., \textit{“Oddly, it appears this building has been newly built up since the disaster”}). Finally, in some cases, participants simply noted that it was difficult to assess building damage, but without necessarily providing a clear reason (E-U; e.g., \textit{“This area is hard to judge”}).

\paragraph{F: Using the number of damaged structures in an image as the measure for severity of the disaster.}
Whereas code D captures cases where participants explain their damage assessments with reference to the apparent extent of damage to the building itself, code F is applied in cases where participants explain their damage assessments with reference to the overall extent of damage observed in the image as a whole. For example, some participants explained their building damage assessment with reference to the number of other buildings that appeared to be affected (F-1; e.g., \textit{“It appears that one building has disappeared, leading me to believe it was destroyed. However, the remaining buildings seen are unharmed”}). Furthermore, other participants explained their damage assessment with reference to the extent of damage visible in the area surrounding a building, either including building damage (F-2; e.g., \textit{“None of the large buildings appear to be damaged, but there is evidence of a large mud patch (in the surrounding area), indicating some minor flood damage”}) or excluding building damage (F-3; e.g., \textit{“All trees have been damaged or destroyed”} and \textit{“Flooding has completely surrounded many of these structures”}). Finally, ambiguous cases were marked as F-U (e.g., \textit{“Every area was totally destroyed”}).

\paragraph{N: No damage \& O: Other minor codes.}

There were some cases where the participant argued that there was ‘no damage,’ without providing further explanation (N). We also identified several other explanation types that arose only once in our dataset. For example, some participants explained their damage assessment with reference to evidence that a specific facility was still operational (e.g., \textit{“Cars are still riding on the road (so the road is not affected)”)}. These explanations were coded as O.

\subsection{Summary of each code}

\begin{figure}[h!]
    \centering
    \includegraphics[width=.8\textwidth]{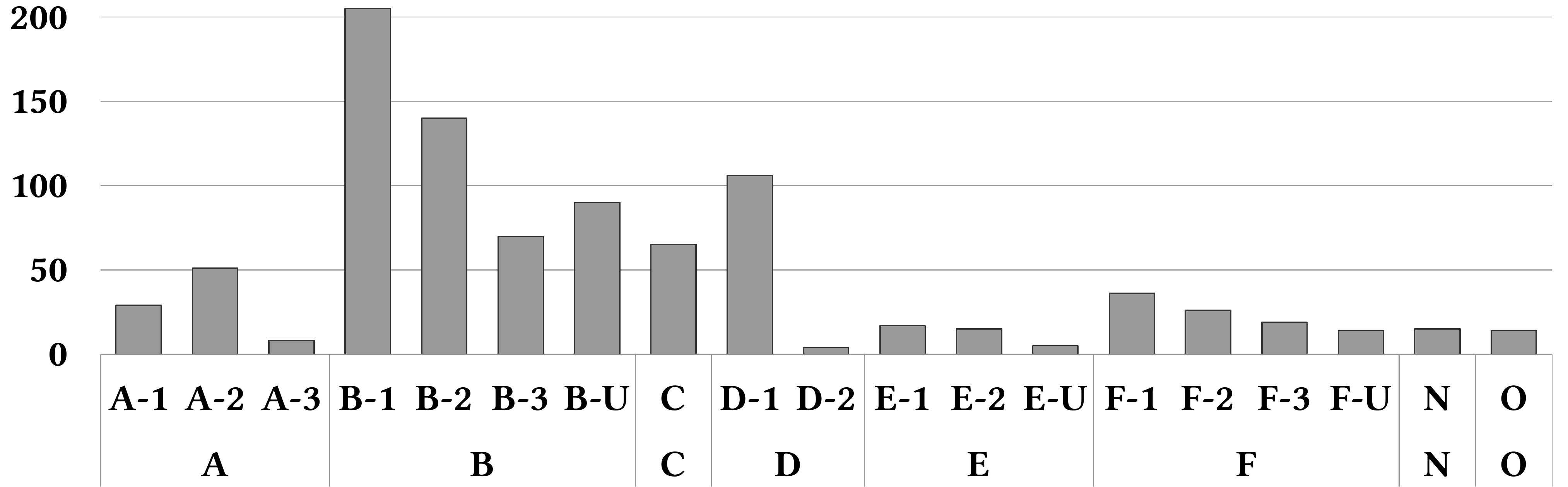}
    \caption{Frequency of each code}
    \label{fig:frequency}
\end{figure}

The frequency of each code is illustrated in Figure~\ref{fig:frequency}. Based on responses from 60 participants, we were able to identify a total of 929 codes (1.55 codes per pre/post image pair on average, \textit{SD} = 1.05). Meanwhile, 12.8\% of the assessments (N = 77) were not assigned to any code, given that participants had not provided explanations for these cases.
\section{Conclusion \& Future Work}

In this work, we analyzed the types of explanations that humans generate when assessing the severity of building damage from satellite imagery. We identified six major strategies that people used (codes A through F), illustrating each with concrete examples.

We observed that, in their explanations, participants often made use of contextual information (e.g., referring to the appearance of the surrounding area or apparent damage to surrounding buildings) in addition to direct evidence of building damage, weaving together these sources of evidence into a coherent story. Participants also frequently made reference to visual contrasts between pre- and post-disaster images, while arguing for particular causal interpretations of these contrasts (e.g., flooding, as in Figure~\ref{fig:examples}-b). Finally, participants sometimes signaled their level of confidence in their own damage assessments within their explanations, while providing reasons for low confidence.

These human explanation strategies stand in contrast to those found in existing AI explanation techniques. It is our hope that the common human explanation strategies we have identified can provide useful inspiration for future AI explanation approaches. However, it remains an open question for future work whether and how these kinds of human explanation strategies might be helpful to consumers of explanations (i.e., HADR decision-makers). Future studies should explore how different types of explanations are perceived by HADR decision-makers, and how different explanation strategies might impact HADR decision-making in practice.

\small
\bibliography{10-reference}

\begin{thebibliography}{8}
\providecommand{\natexlab}[1]{#1}
\providecommand{\url}[1]{\texttt{#1}}
\expandafter\ifx\csname urlstyle\endcsname\relax
  \providecommand{\doi}[1]{doi: #1}\else
  \providecommand{\doi}{doi: \begingroup \urlstyle{rm}\Url}\fi

\bibitem[{Barredo Arrieta} et~al.(2020){Barredo Arrieta}, Díaz-Rodríguez,
  {Del Ser}, Bennetot, Tabik, Barbado, Garcia, Gil-Lopez, Molina, Benjamins,
  Chatila, and Herrera]{arrieta2020explainable}
Alejandro {Barredo Arrieta}, Natalia Díaz-Rodríguez, Javier {Del Ser}, Adrien
  Bennetot, Siham Tabik, Alberto Barbado, Salvador Garcia, Sergio Gil-Lopez,
  Daniel Molina, Richard Benjamins, Raja Chatila, and Francisco Herrera.
\newblock {Explainable Artificial Intelligence (XAI): Concepts, taxonomies,
  opportunities and challenges toward responsible AI}.
\newblock \emph{Information Fusion}, 58:\penalty0 82--115, 2020.

\bibitem[Durnov(2019)]{xview_first}
Victor Durnov.
\newblock {xView2 first place}, 2019.
\newblock URL \url{https://github.com/DIUx-xView/xView2_first_place}.

\bibitem[Gupta et~al.(2019)Gupta, Hosfelt, Sajeev, Patel, Goodman, Doshi, Heim,
  Choset, and Gaston]{gupta2019xbd}
Ritwik Gupta, Richard Hosfelt, Sandra Sajeev, Nirav Patel, Bryce Goodman, Jigar
  Doshi, Eric Heim, Howie Choset, and Matthew Gaston.
\newblock {xBD: A Dataset for Assessing Building Damage from Satellite
  Imagery}.
\newblock \emph{arXiv preprint arXiv:1911.09296}, 2019.

\bibitem[Logar et~al.(2020)Logar, Bullock, Nemni, Bromley, Quinn, and
  Luengo-Oroz]{logar2020pulsesatellite}
Tomaz Logar, Joseph Bullock, Edoardo Nemni, Lars Bromley, John~A Quinn, and
  Miguel Luengo-Oroz.
\newblock {PulseSatellite: A tool using human-AI feedback loops for satellite
  image analysis in humanitarian contexts}.
\newblock In \emph{Proceedings of the AAAI Conference on Artificial
  Intelligence}, volume~34, pages 13628--13629, 2020.

\bibitem[Miller(2019)]{miller2019explanation}
Tim Miller.
\newblock {Explanation in artificial intelligence: Insights from the social
  sciences}.
\newblock \emph{Artificial Intelligence}, 267:\penalty0 1--38, 2019.

\bibitem[Mohseni et~al.(2021)Mohseni, Zarei, and
  Ragan]{mohseni2021multidisciplinary}
Sina Mohseni, Niloofar Zarei, and Eric~D Ragan.
\newblock {A multidisciplinary survey and framework for design and evaluation
  of explainable AI systems}.
\newblock \emph{ACM Transactions on Interactive Intelligent Systems (TiiS)},
  11\penalty0 (3-4):\penalty0 1--45, 2021.

\bibitem[Walker and Myrick(2006)]{walker2006grounded}
Diane Walker and Florence Myrick.
\newblock {Grounded theory: An exploration of process and procedure}.
\newblock \emph{Qualitative health research}, 16\penalty0 (4):\penalty0
  547--559, 2006.

\bibitem[Zhang et~al.(2019)Zhang, Zhang, Li, Plummer, and
  Wang]{zhang2019crowdlearn}
Daniel Zhang, Yang Zhang, Qi~Li, Thomas Plummer, and Dong Wang.
\newblock {Crowdlearn: A crowd-ai hybrid system for deep learning-based damage
  assessment applications}.
\newblock In \emph{2019 IEEE 39th International Conference on Distributed
  Computing Systems (ICDCS)}, pages 1221--1232. IEEE, 2019.

\end{thebibliography}

\end{document}